\documentclass[%
 reprint,
 superscriptaddress,
 amsmath,amssymb,
 aps,
prb,
]{revtex4-2}
\usepackage{graphicx}
\usepackage{subfigure}
\usepackage{float}
\usepackage{dcolumn}
\usepackage{bm}
\usepackage{rotating}
\usepackage{color}
\usepackage{xcolor}
\usepackage{braket}

\usepackage[colorlinks,citecolor=blue,urlcolor=blue,linkcolor=blue]{hyperref}
\begin{document}

\preprint{APS/123-QED}


\title{Investigation of reentrant localization transition in one-dimensional quasi-periodic lattice with long-range hopping}

\author{Pei-Jie Chang}
    \affiliation{State Key Laboratory of Low-Dimensional Quantum Physics and Department of Physics, Tsinghua University, Beijing 100084, China}
    
\author{Qi-Bo Zeng} 
	\affiliation{Department of Physics, Capital Normal University, Beijing 100048, China}

\author{Jinghui Pi}
 	\affiliation{The Chinese University of Hong Kong Shenzhen Research Institute, 518057 Shenzhen, China}
	\affiliation{Department of Physics, The Chinese University of Hong Kong, Shatin, New Territories, Hong Kong, China}

\author{Dong Ruan}
    \affiliation{State Key Laboratory of Low-Dimensional Quantum Physics and Department of Physics, Tsinghua University, Beijing 100084, China}
    \affiliation{Frontier Science Center for Quantum Information, Tsinghua University, Beijing 100084, China}

\author{Gui-Lu Long}
    \email{gllong@tsinghua.edu.cn}
    \affiliation{State Key Laboratory of Low-Dimensional Quantum Physics and Department of Physics, Tsinghua University, Beijing 100084, China}
    \affiliation{Frontier Science Center for Quantum Information, Tsinghua University, Beijing 100084, China}
    \affiliation{Beijing Academy of Quantum Information Sciences, Beijing 100193, China}
    \affiliation{Beijing National Research Center for Information Science and Technology, Beijing 100084, China}

\begin{abstract}

Reentrant localization has recently been observed in systems with quasi-periodic nearest-neighbor hopping, where the interplay between dimerized hopping and staggered disorder is identified as the driving mechanism. However, the robustness of reentrant localization in the presence of long-range hopping remains an open question. In this work, we investigate the phenomenon of reentrant localization in systems incorporating long-range hopping. Our results reveal that long-range hopping induces reentrant localization regardless of whether the disorder is staggered or uniform. We demonstrate that long-range hopping does not inherently disrupt localization; instead, under specific conditions, it facilitates the emergence of reentrant localization. Furthermore, by analyzing critical exponents, we show that the inclusion of long-range hopping modifies the critical behavior, leading to transitions that belong to distinct universality classes. 

\end{abstract}

\maketitle


\section{Introduction}

Anderson localization is one of the key topics in condensed matter physics \cite{PhysRev.109.1492, RevModPhys.57.287, RevModPhys.80.1355}. It refers to the suppression of wave propagation, such as electronic transport or light diffusion, in disordered media due to interference effects arising from multiple scattering. This phenomenon has been extensively studied in a variety of physical systems and has been experimentally verified in numerous works \cite{roati2008anderson, PhysRevLett.103.013901, aspect2009anderson, kondov2011three, jendrzejewski2012three, semeghini2015measurement, PhysRevLett.118.170403, hainaut2019experimental, PhysRevLett.122.100403, maynard2001acoustical, Transversal, hu2008localization, yamilov2023anderson, segev2013anderson, mookherjea2014electronic, vatnik2017anderson, PhysRevLett.100.013906, kovacs2010anderson, ghasempour2023anderson, crespi2013anderson, chen2014emulating, PhysRevA.89.022309}. According to scaling theory \cite{PhysRevLett.42.673}, all single-particle states in one-dimensional (1D) and two-dimensional systems with uncorrelated disorder exhibit localization. In three-dimensional systems, increasing the strength of uncorrelated disorder gives rise to a mobility edge, which defines the boundary between localized and extended states. However, in systems with correlated or structured disorder, deviations from scaling theory can occur, allowing for disorder induced metal-insulator transitions in low-dimensional systems. The Aubry-André-Harper (AAH) model provides a framework for studying the impact of quasi-periodic disorder on localization in 1D systems \cite{AAH, harper1955single}, which introduces a potential with an incommensurate period relative to the lattice sites. Due to its self-duality, the AAH model predicts that all eigenstates undergo a sharp transition from extended to localized as the system crosses the critical point, which has been confirmed across various physical realizations \cite{roati2008anderson, PhysRevLett.103.013901, PhysRevLett.109.106402, xue2014observation}. Generalizations of the AAH model, such as those incorporating power-law distributed long-range hopping \cite{kim2010quantum, richerme2014non, britton2012engineered, islam2013emergence, schauss2015crystallization, PhysRevLett.121.093602, PhysRevB.103.075124, roy2021entanglement, PhysRevB.110.094203, PhysRevResearch.3.013148} or mosaic-type disorder \cite{PhysRevLett.125.196604, PhysRevB.108.054204, PhysRevB.104.064203}, reveal the emergence of mobility edges in 1D systems \cite{PhysRevLett.104.070601, PhysRevLett.114.146601, PhysRevLett.123.070405}. These models provide valuable opportunities to study the properties of mobility edges in low-dimensional systems, offering deeper insights into the nature of localization transitions and the influence of different types of disorder on system behavior.

Recent studies have revealed that in 1D quasi-periodic Su–Schrieffer–Heeger (SSH) systems, increasing the quasi-periodic disorder induces a reentrant localization transition \cite{PhysRevLett.126.106803}. Specifically, for certain dimerization strengths, some eigenstates of the system undergo a sequence of transitions: initially localized, then delocalized, and eventually localized again as the disorder strength increases. This intriguing phenomenon has garnered significant attention \cite{PhysRevB.106.054307, PhysRevB.105.054204, PhysRevB.108.L100201, PhysRevB.105.L220201, PhysRevA.106.013305, wu2021non, PhysRevB.110.134203, PhysRevB.107.224201, PhysRevB.107.035402, ganguly2024phenomenon, li2023multiple, PhysRevA.106.053312, PhysRevA.109.043308, lu2023robust, PhysRevB.109.195427, PhysRevB.110.184208, sarkar2024signature, gong2021comment}. Subsequent investigations have demonstrated that reentrant localization transitions, along with the associated mobility edges, can persist in non-Hermitian systems \cite{PhysRevB.105.054204, PhysRevB.109.L020203, PhysRevA.108.033305, jiang2021mobility} or systems with specific long-range hopping strengths \cite{PhysRevB.107.075128}. For instance, experimental studies on $Si_3N_4$ waveguide systems have introduced random-dimer disorder, in contrast to quasi-periodic disorder, and observe analogous reentrant localization transitions \cite{xu2024observation}. Similar phenomena have also been reported in photonic crystal experimental setups \cite{PhysRevResearch.5.033170}. Furthermore, research into the critical exponents of localization transitions has revealed a notable consistency: the critical exponents at the second and third transition points in these systems are identical, suggesting that these transitions belong to the same universality class \cite{PhysRevB.105.214203}. These findings deepen our understanding of localization phenomena and the conditions under which reentrant behavior arises in various physical systems. 

In this study, we investigate the 1D SSH quasi-periodic model with long-range hopping and examine localization transitions under both staggered and uniform disorder conditions. We show that the inclusion of long-range hopping not only mitigates the effect of reentrant localization but can also induce reentrant localization under specific conditions. For staggered disorder, we observe a reentrant localization phenomenon by analyzing the phase diagram and the spatial distribution of eigenstates across lattice sites. Furthermore, the system exhibits the simultaneous presence of multiple pairs of mobility edges. A detailed analysis of the critical exponents at various phase transition points reveals that long-range hopping alters the nature of localization transitions, placing them in distinct universality classes. In the case of uniform disorder, reentrant localization and multiple pairs of mobility edges are observed within the same numerical range through phase diagram analysis. Similar to the staggered disorder case, calculations of the critical exponents at different transition points demonstrate that long-range hopping leads to transitions belonging to different universality classes. 

The remainder of the paper is organized as follows: In Sec. \ref{se2}, we provide a detailed description of the model under consideration and the computational approach employed. In Sec. \ref{se3} and Sec. \ref{se4}, we present our main findings for systems with staggered disorder and uniform disorder, respectively. These sections focus on the confirmation of reentrant localization and the analysis of critical exponents. And in Sec. \ref{se5}, we explore the effects of inter-leg hopping on reentrant localization. Finally, a summary of our conclusions is provided in Sec. \ref{se6}. 

\section{MODEL AND APPROACH}\label{se2}

\begin{figure}
     \raggedright
     \includegraphics[width=0.5\textwidth]{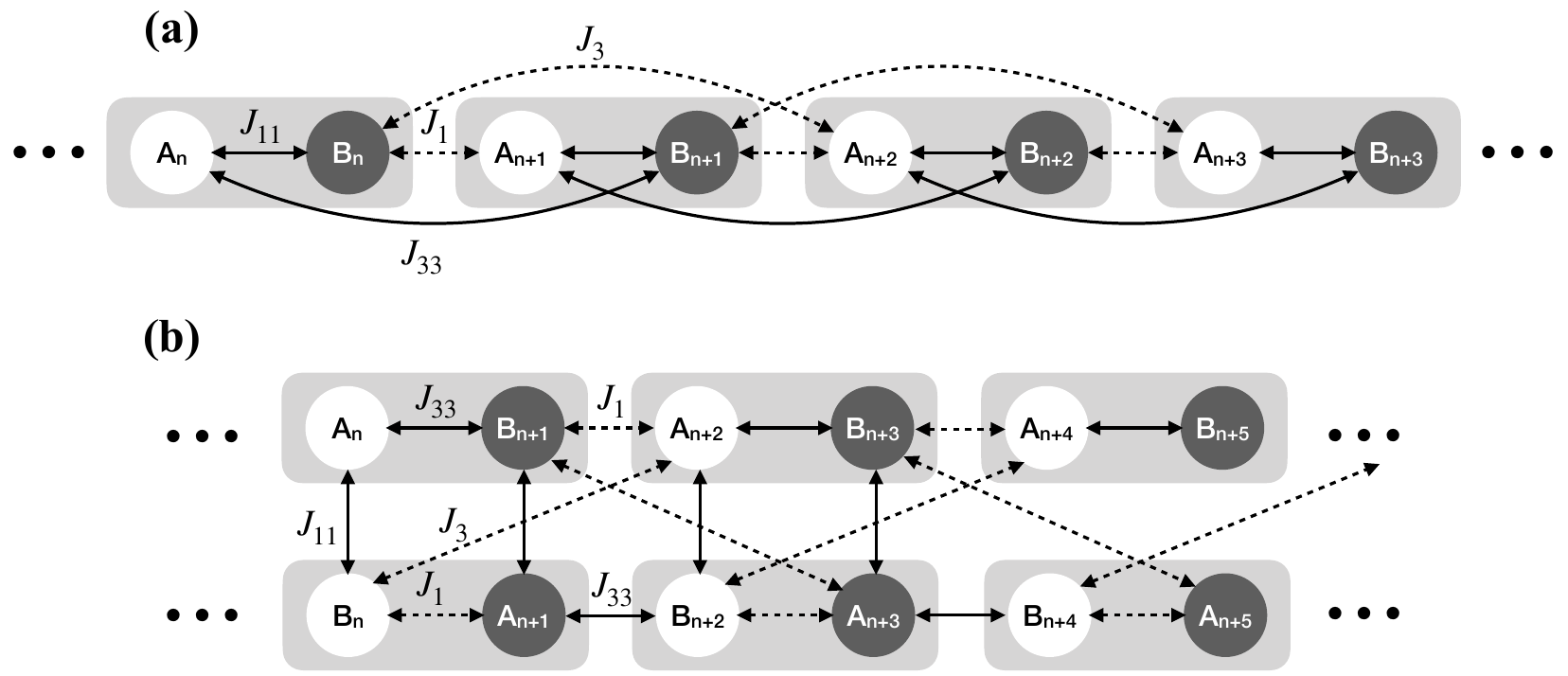}
     \caption{(a) Schematic representation of the modulated SSH model, incorporating next-nearest-neighbor hopping terms and a quasi-periodic potential. This model consists of two sublattices, $A$ and $B$. $J_1$, $J_{11}$, $J_{3}$, $J_{33}$ represent inter-cell hopping, intra-cell hopping and the next-nearest hopping with different sub-lattices, respectively. (b) Schematic representation of the 1D model reformulated as a ladder-like structure. }
     \label{fig0}
\end{figure}

We consider a SSH model that includes next-nearest-neighbor hopping and incorporates either a staggered or uniform quasi-periodic disorder term. The Hamiltonian of this model is expressed as
\begin{equation}
    H=H_0+H_d,
    \label{model}
\end{equation}
with hopping term
\begin{equation}
    \begin{aligned}
        H_{0}= &  -\sum_{i=0}^{N}(J_{11}\hat{c}_{i,A}^{\dagger}\hat
        {c}_{i,B}+J_{1}\hat{c}_{i,B}^{\dagger}\hat{c}_{i+1,A}
        +J_{33}\hat{c}_{i,A}^{\dagger}\hat{c}_{i+1,B}\\
        &  +J_{3}\hat{c}_{i,B}^{\dagger}\hat{c}_{i+2, A}+h.c.)\newline,
    \end{aligned}
\end{equation}
and the onsite quasi-periodic disorder term
\begin{equation}
	\begin{aligned}
		H_d = & \sum_{i=0}^{N} \lambda_A\hat{n}_{i, A}cos\left[ 2\pi\beta(2i-1)+\theta\right] \\
		&+\sum_{i=0}^{N} \lambda_B\hat{n}_{i, B}cos\left[2\pi\beta(2i)+\theta\right],  
	\end{aligned}
\end{equation}
where with $L=2N$ representing the total number of lattice sites. Following the standard approach of the SSH model, the lattice is divided into two sub-lattices, $A$ and $B$, with the corresponding creation and annihilation operators defined as: $\hat{c}_{i, A}^{\dagger}(\hat{c}_{i, A})$ and $\hat{c}_{i, B}^{\dagger}(\hat{c}_{i, B})$, respectively. Operators $\hat{n}_{i,A}$ and $\hat{n}_{i,B}$ represent the particle number operators for the corresponding sub-lattices $A$ and $B$, respectively. The parameters $J_1$ and $J_{11}$ denote the inter-cell and intra-cell hopping amplitudes for nearest neighbour, respectively. And $J_3$ and $J_{33}$ correspond to the next-nearest neighbour hopping amplitudes between different sub-lattices A and B as shown in Fig.~\ref{fig0}(a). For convenience, we set $J_1$ as the unit of the energy scale throughout our study. The on-site quasi-periodic potential differs for the two sublattices: $\lambda_A$ and $\lambda_B$ represent the potential strengths for sublattices $A$ and $B$, respectively. The quasi-periodic potential is implemented by selecting $\beta$ as a Diophantine number to ensure incommensurability. To minimize finite-size effects, we perform simulations on systems with sizes up to $35422$ sites. Additionally, following standard practice in the literature and without loss of generality, we set $\theta=0$. 

We diagonalize the Hamiltonian in the single-particle lattice representation, which allows us to obtain the eigenenergy $E_m$ and the corresponding eigenstates of Eq. \ref{model} as 
\begin{equation}
	\ket{\psi_m} = \sum_{i}^{L} \phi_i^{(m)}\ket{i}.
\end{equation}
Here, $\phi_i^{(m)}$ represents the probability amplitude of the corresponding $m_{st}$ eigenstate at $i_{st}$ site. The localization properties of the system, which aimed to study, can be fully characterized by these eigenstates. To distinguish whether the eigenstates are localized, critical, or extended, we calculate the inverse participation ratio (IPR) and the normalized participation ratio (NPR) of the system. For the $m_{th}$ eigenstate $\phi^{(m)}$, the definitions of the IPR and NPR are as follows:
\begin{equation}
	\begin{aligned}
		& IPR_{(m)} = \sum_{i=1}^{L}|\phi_i^{(m)}|^4, \\
		& NPR_{(m)} = \left(L\sum_{i=1}^{L}|\phi_i^{(m)}|^4\right)^{-1}.
	\end{aligned}
\end{equation}
For extended states, the IPR scales as $1/L$, approaching 0 as the system size $L$ becomes large, while the NPR remains nonzero. For localized states, the NPR scales as $1/L$, tending to 0 as the system size $L$ increases, while the IPR remains nonzero. For critical states, the system exhibits eigenstates where both the IPR and NPR are finite values simultaneously. To more conveniently track how the system’s states evolve with varying parameters, we calculate the average IPR and NPR within a specific energy interval, defined as 
\begin{equation}
	\begin{aligned}
		& IPR = \frac{1}{m}\sum_{m}IPR_{(m)}, \\
		& NPR = \frac{1}{m}\sum_{m}NPR_{(m)}.
	\end{aligned}
	\label{IPRNPR}
\end{equation}
For the subsequent analysis of the critical exponents, we define the mean square NPR as
\begin{equation}
	\sigma=\sqrt{NPR}. 
\end{equation}

Near the phase transition critical point, the critical behavior of a parameter can be described by the following power law:
\begin{equation}
	\sigma\sim(-\varepsilon)^{\beta},\ L*NPR\sim\varepsilon^{-\gamma},\ \xi\sim|\varepsilon|^{-\nu}.  
	\label{critical}
\end{equation}
Here, $\varepsilon=(\lambda-\lambda_c)/\lambda_c$ with $\lambda_c$ is the critical quasi-periodic disorder strength for the localization transition, and $\xi$ represents the correlation length. According to \cite{hashimoto1992finite, PhysRevB.105.214203}, we can calculate the $R$ function for systems of different sizes
\begin{equation}
	R[L, L^{'}]=\frac{ln(\sigma_L^2/\sigma_{L^{'}}^2)}{ln(L/L^{'})}+1. 
	\label{R equ}
\end{equation}
The curves intersect at a single common point at the localization transition for different system sizes $L$ and $L^{'}$. The horizontal coordinate of the intersection point corresponds to the critical disorder strength $\lambda_c$, while the vertical coordinate gives the ratio of critical exponents $\gamma/\nu$. Near the localization transition point, systems of different size $L$ follow the same scaling function $G$ as 
\begin{equation}
	\sigma^2=L^{\gamma/\nu-1}G(\varepsilon L^{1/\nu}).
	\label{nu}
\end{equation}
Through this relationship, the critical exponent $\nu$ can be determined by minimizing the relative error of scaling function $G$, which allows for the analysis of the critical properties near different localization transition critical points.

As shown in \cite{PhysRevLett.126.106803}, it is found that under staggered disorder, defined as $\lambda_A=-\lambda_B$, the system exhibits a reentrant localization transition. Calculations of the critical exponents indicate that the second and third transitions belong to the same universality class. In contrast, no reentrant localization transition is observed under uniform disorder, defined as $\lambda_A=\lambda_B$. However, in systems with long-range hopping, both staggered disorder and uniform disorder display reentrant localization features in the phase diagram, with critical exponents differing from those observed in systems without long-range hopping. 

In the next section, we focus on the discussing the influence of long range hopping under staggered disorder and uniform disorder separately. To better characterize the localization transitions between the critical state and other states, we define quantity $\eta$ consistent with \cite{PhysRevLett.126.106803, PhysRevB.101.064203} as follows:
\begin{equation}
	\eta=log_{10}(IPR\times NPR), 
\end{equation}
where IPR and NPR are taken the average over all the eigenstates as shown in Eq. \ref{IPRNPR}. A smaller value of $\eta$ corresponds to an extended or localized state, while a larger value of $\eta$ indicates a critical state.

\begin{figure}
     \centering
     \includegraphics[width=0.5\textwidth]{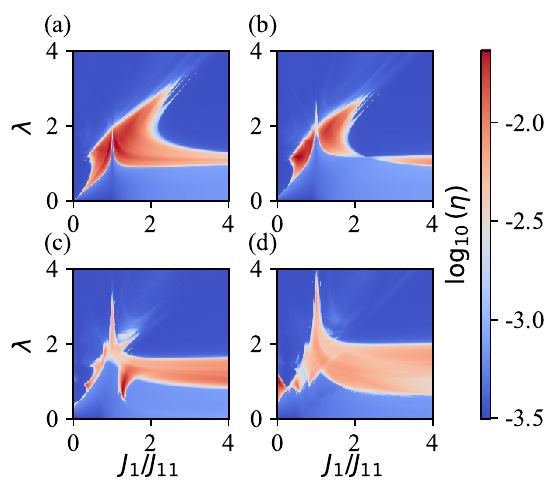}
     \caption{$\eta$ phase diagrams of the system in $\lambda$ and $J_1/J_{11}$ plane with $J_3=0$ for (a) $J_{33}=0$, (b) $J_{33}=0.1$, (c) $J_{33}=0.3$, (d) $J_{33}=0.5$, where the system size $L=3194$ and $\lambda_A=-\lambda_B$. The color represents different values of $\log_{10}(\eta)$. And the red region corresponds to the intermediate states where single-particle mobility edges exist, while the blue region indicates that the system is entirely in the localized or extended state. }
     \label{fig1}
\end{figure}

\section{Staggered disorder}\label{se3}

\begin{figure}
    \centering
    \includegraphics[width=0.5\textwidth]{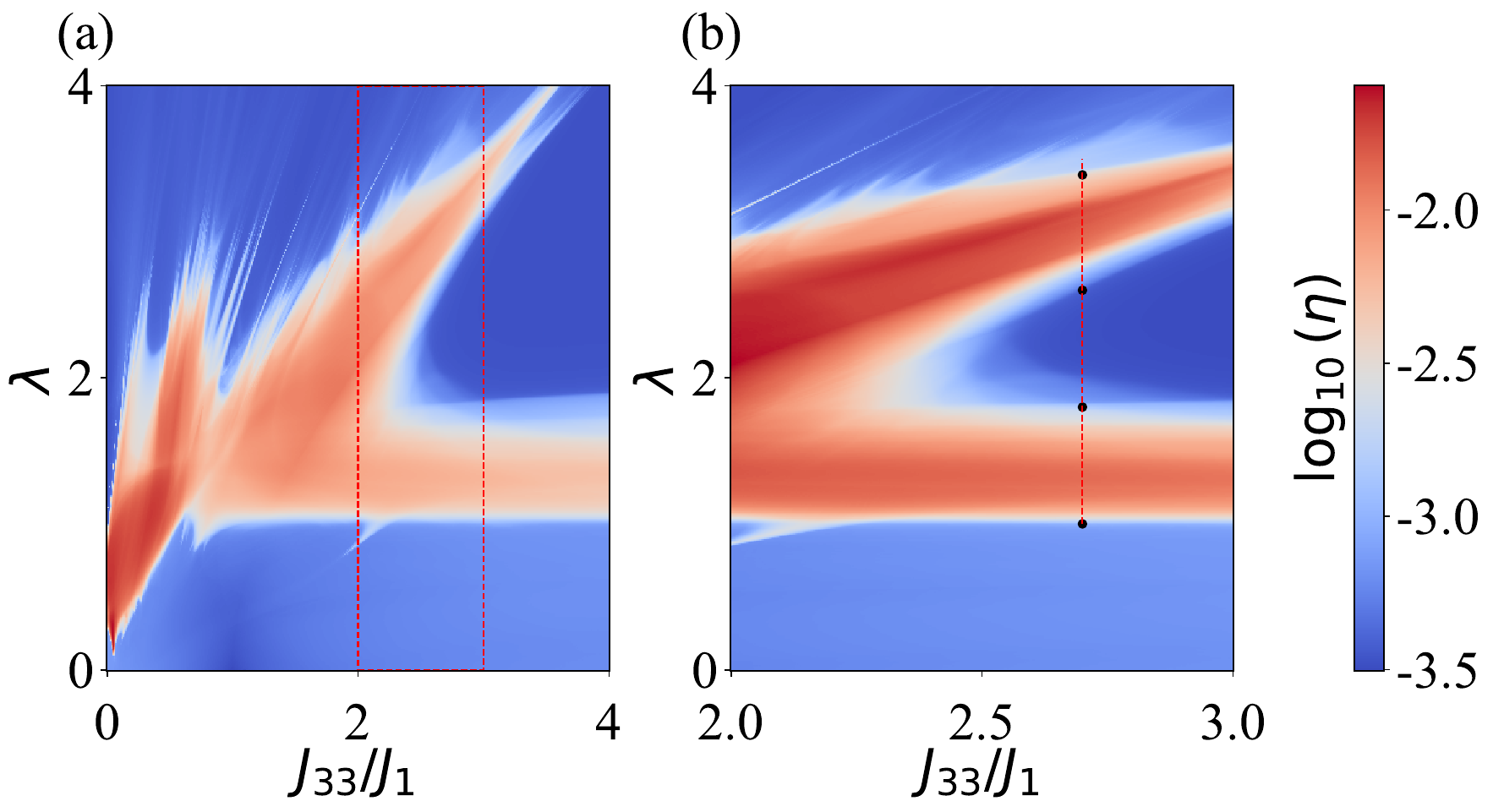}
    \caption{$\eta$ phase diagrams of the system in $\lambda$ and $J_{33}/J_{1}$ plane with $J_{11}=J_3=0.1$, the system size $L=3194$, $\beta=\frac{\sqrt{7}-1}{2}$ and $\lambda_A=-\lambda_B$. The color represents different values of $\log_{10}(\eta)$. (b) a zoomed-in version of the portion enclosed by the dashed box in (a). The dashed line in (b) is located at $J_{33}/J_{1}=2.7$, and the black dots represent schematic localization transition points. }
    \label{fig2}
\end{figure}

\begin{figure}
     \raggedright
     \includegraphics[width=0.48\textwidth]{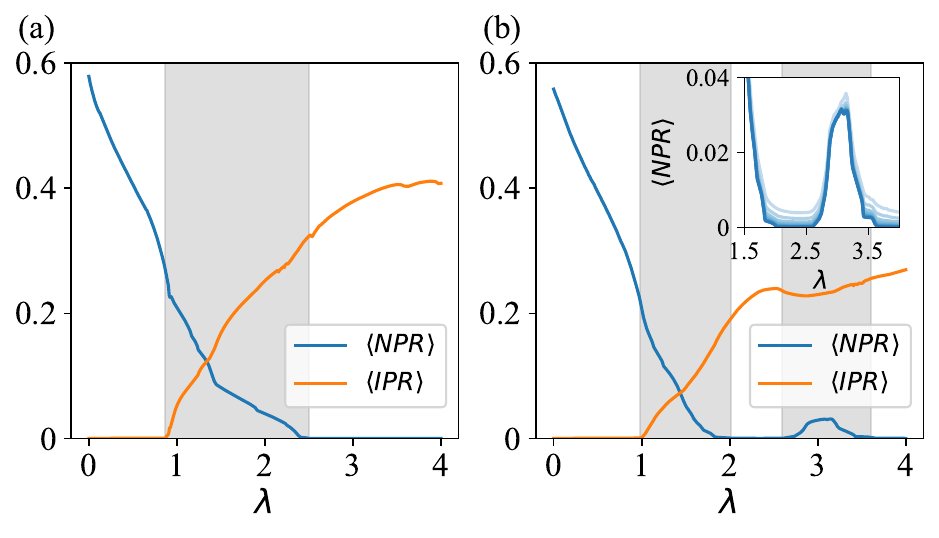}
     \caption{(a) and (b) show the average IPR and NPR over all eigenstates for $J_{33}/J_1=0.5$ and $J_{33}/J_1=2.7$, respectively, for the case of staggered disorder and system size $L=13 530$. The shaded regions represent the critical area, where both localized and extended states coexist. The inset in (b) displays the average NPR for $L=1220,  1974,  3194,  5168,  8362, 13530, 21892,$ and $35422$, with the color intensity ranging from light to dark. }
     \label{fig3}
\end{figure}

\begin{figure}
     \centering
     \includegraphics[width=0.48\textwidth]{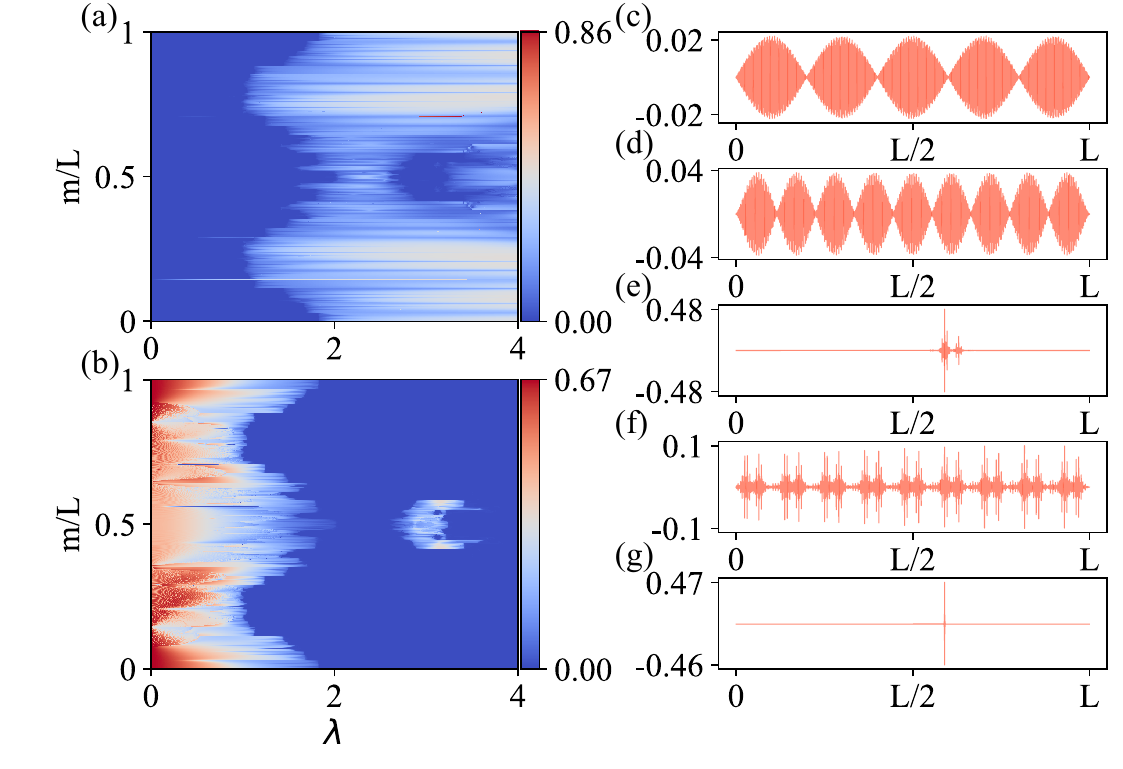}
     \caption{(a) The IPR and (b) NPR associated with the eigenstate indices as a function of $\lambda$ for $J_{33}/J_1=2.7$, respectively. (c) to (g) illustrate the spatial distribution of eigenstates across lattice sites for $\lambda=0.4, 1.2, 2.1, 2.7, 4$, with eigenstate index $m/L=0.5$ and $L=13530$. }
     \label{fig4}
\end{figure}

\begin{figure*}
     \centering
     \includegraphics[width=\textwidth]{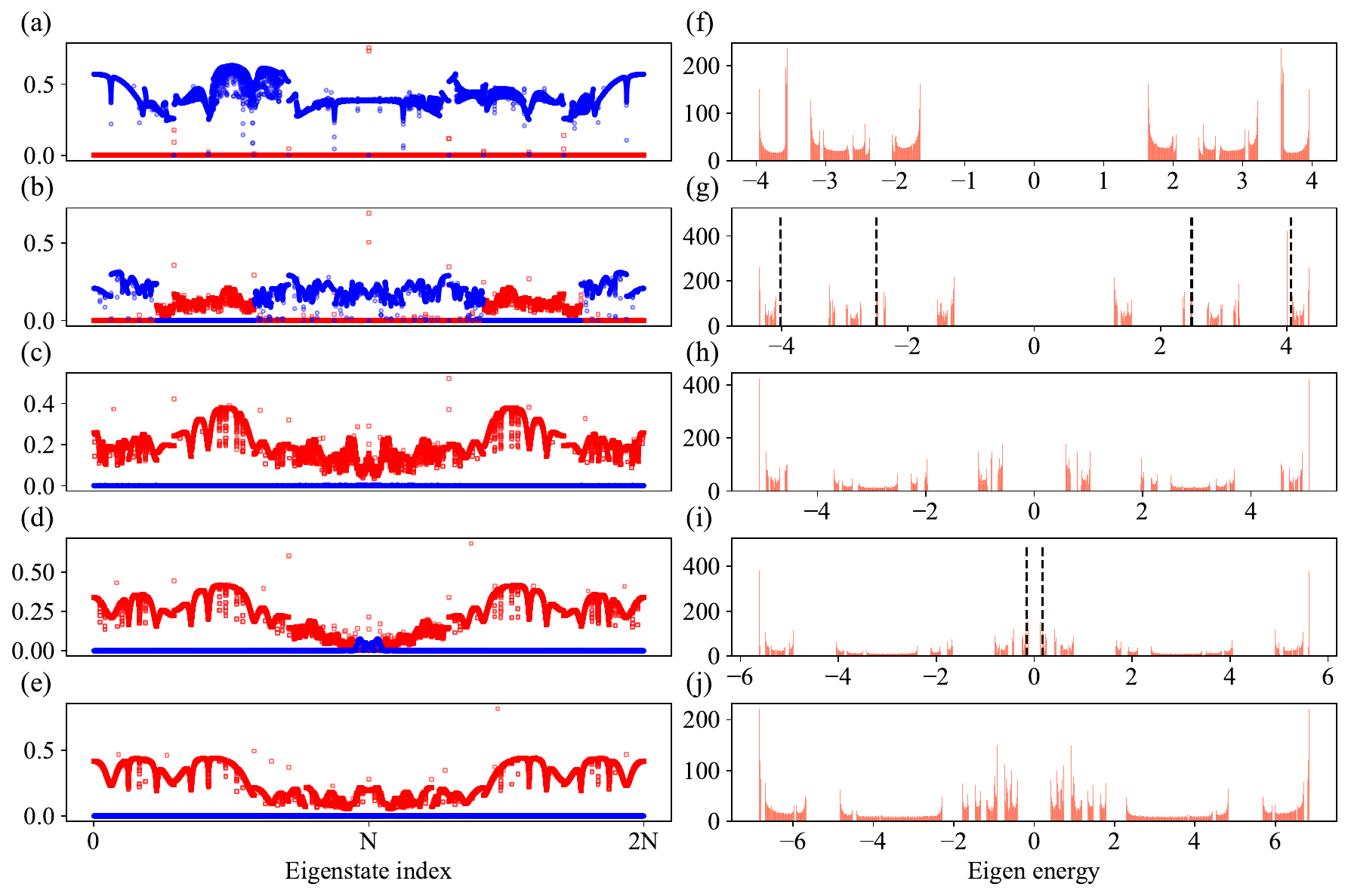}
     \caption{(a)–(e) show the distributions of IPR (red) and NPR (blue) as functions of the eigenstate index for $\lambda=0.4, 1.2, 2.1, 2.7$ and $4$, respectively. (f)–(j) display the density of states (DOS) as a function of eigenenergy for the same $\lambda$ values, with the black dashed lines indicating the locations of single-particle mobility edges. }
     \label{fig5}
\end{figure*}

\begin{figure}
     \raggedright
     \includegraphics[width=0.48\textwidth]{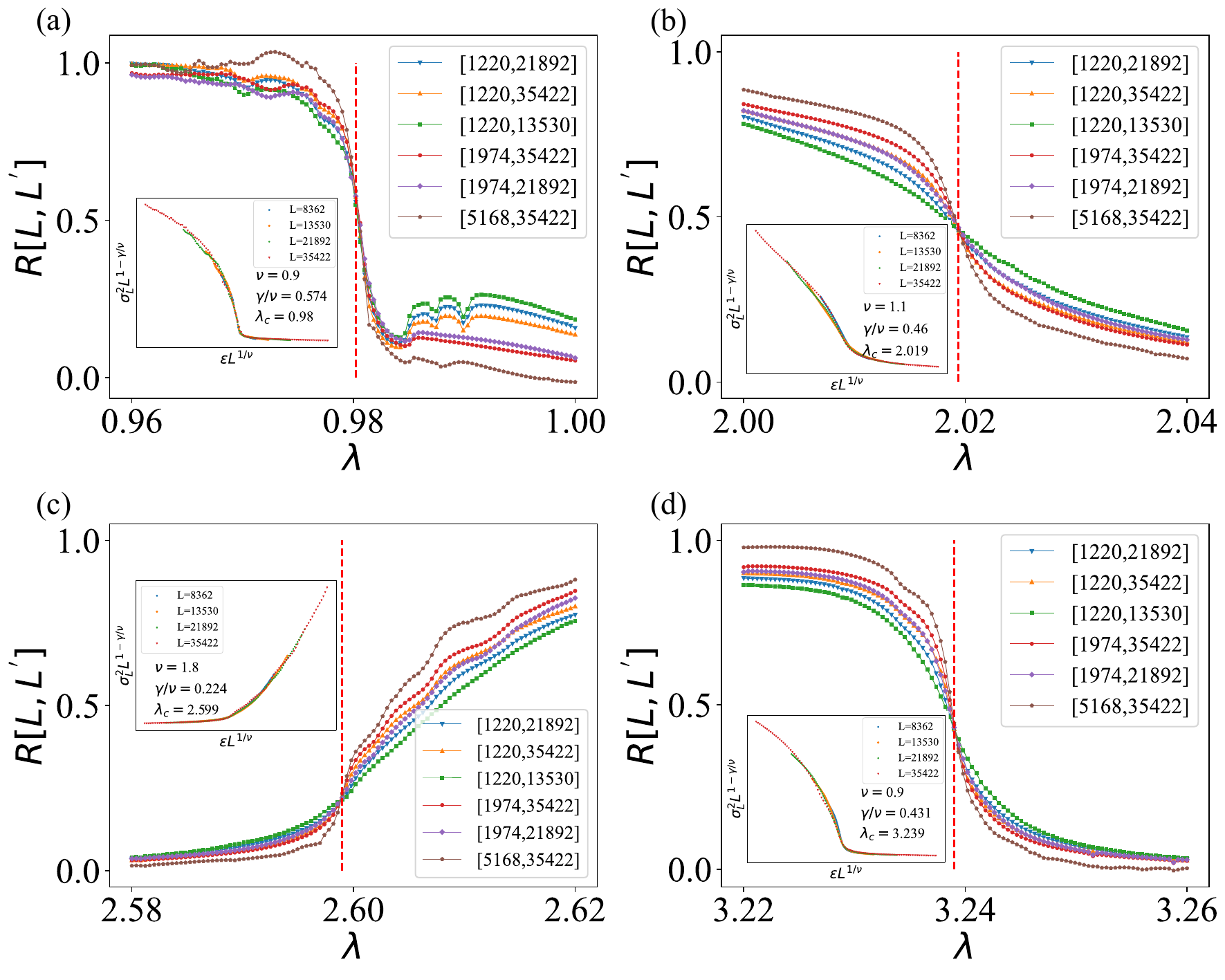}
     \caption{(a)–(d) depict the values of the function $R[L,\ L^{'}]$ near the first, second, third, and fourth critical quasi-periodic disorder potential strength, with $J_{33}/J_1=2.7$ and $\lambda_A=-\lambda_B$. In each panel, the inset presents the $\sigma^2$ of the fitted critical exponent $\nu$ near the respective critical quasi-periodic disorder strength. The coordinate of the function $R[L,\ L^{'}]$ crossing point is denoted as $(\lambda_c, \gamma/\nu)$. For the first critical point, we average the NPR of eigenstates with $m/L \in [0.238, 0.243]$. For the second, third, and fourth critical points, we average the NPR of eigenstates with $m/L \in [0.48, 0.52]$. }
     \label{fig6}
\end{figure}

\begin{figure}
     \raggedright
     \includegraphics[width=0.48\textwidth]{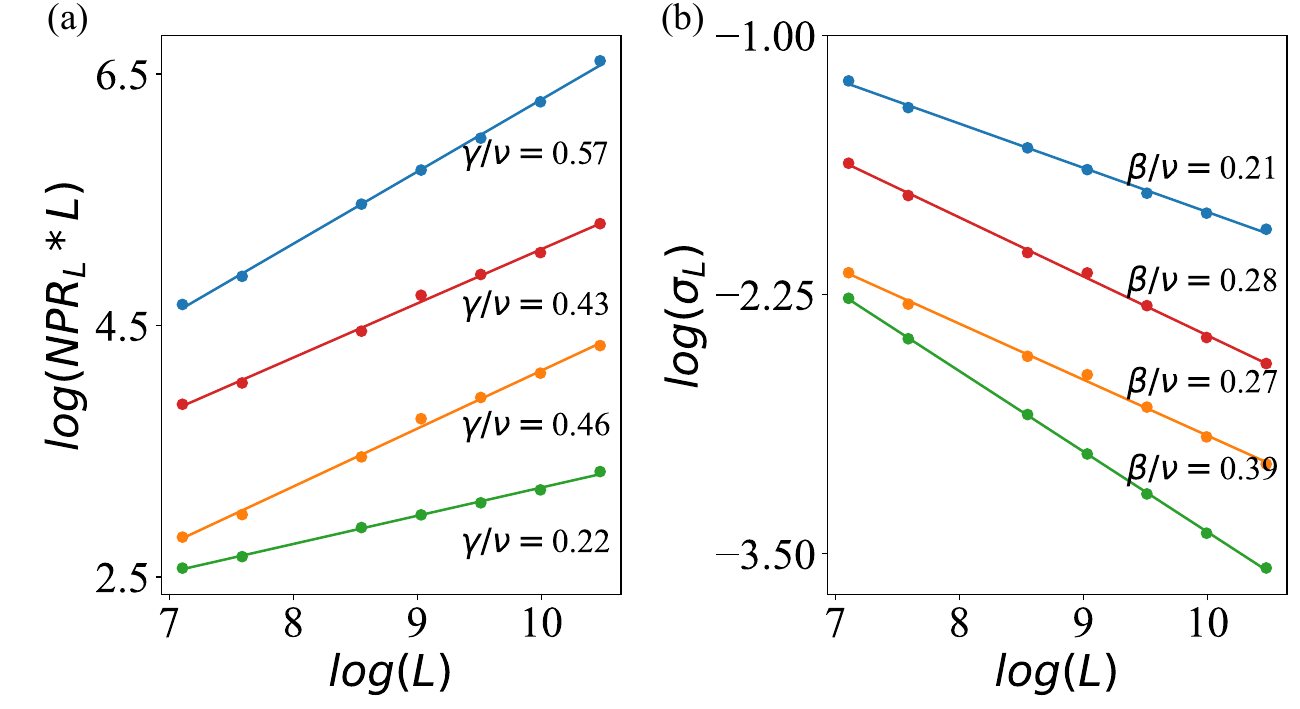}
     \caption{The critical exponent ratio (a) $\gamma/\nu$ and (b) $\gamma/\beta$ for the four critical quasi-potential points. These ratios are obtained by fitting $\log(NPR_L*L)$ against $\log(L)$. The system sizes $L$ used for the fitting are: $1220, 1974, 5168, 8362, 13530, 21892$, and $35422$. }
     \label{fig7}
\end{figure}

Previous literature \cite{PhysRevLett.126.106803} has discussed the occurrence of a reentrant localization transition under staggered disorder. When long-range next-nearest neighbour hopping is simply added to the model with the same parameters, it is observed that the reentrant localization transition gradually weakens and eventually disappears, as shown in Fig.~\ref{fig1}. Specifically, for $L=3194$, $J_3=0$ and varying $J_{33}=0,0.1,0.3,0.5$, the reentrant localization transition essentially disappears when $J_{33}$ reaches approximately $0.3$. This result is consistent with the findings in \cite{PhysRevB.107.075128}, where the introduction of long-range hopping between the same lattice sites in different unit cells weakens the formation of dimers. As a result, the competition between the dimers and disorder diminishes, ultimately preventing the transition from localized to extended states.

However, describing the long-range hopping as merely weakening the reentrant localization transition is not entirely accurate. When the system is tuned to appropriate parameters, such as $J_{11}=J_3=0.1$, the reentrant localization transition still appears in the $\eta$ phase diagram for the same system size, as clearly demonstrated in Fig.~\ref{fig2}. We should point out that this observation does not completely contradict previous conclusions of \cite{PhysRevB.107.075128}. In the presence of long-range hopping, the originally 1D chain-like system can be effectively reduced to a 1D ladder structure as shown in Fig~\ref{fig0}(b). In this ladder configuration, the emergence of reentrant localization can be understood as a competition between dimers and disorder terms in each leg, similar to the 1D chain, which leads to the reentrant localization transition. Therefore, even with long-range hopping, the characteristic of reentrant localization can still arise under conditions of small inter-leg hopping. 

Moreover, the behavior of reentrant localization in this case is not directly related to the topological phase transition of the system. As shown in Fig.~\ref{fig2}, the reentrant localization feature begins to emerge when $J_{33}/J_1>2.3$. In the presence of long-range hopping, the SSH model exhibits a multi-phase diagram, with the phase transition occurring near $J_{33}=J_1$ \cite{perez2018ssh, PhysRevB.99.035146}. This suggests that the emergence of reentrant localization is not a result of the topological phase transition even with long-range hopping. 

In Fig.~\ref{fig3}(a) and~\ref{fig3}(b), we present the average IPR and NPR over all eigenstates for two different cases, $J_{33}/J_1 = 0.5$ and $J_{33}/J_1 = 2.7$, as marked by the dotted line in~\ref{fig2}(b). As shown in Fig.~\ref{fig3}(a) for $J_{33}/J_1 = 0.5$, a critical region appears when $0.86 < \lambda < 2.5$. After the localization transition, for $\lambda > 2.5$, all eigenstates of the system remain localized, and no further extended states appear as $\lambda$ increases. In contrast, for $J_{33}/J_1 = 2.7$ in Fig.~\ref{fig3}(b), two critical regions are observed: $0.98 < \lambda < 2.018$ and $2.599 < \lambda < 3.6$, where both IPR and NPR take finite values. Between these two critical regions and beyond the second critical region, all eigenstates exhibit localized characteristics. It is noteworthy that the second critical region is much smaller than the first. To rule out the possibility of finite-size effect, we calculate the results for different system sizes: $L=1220, 1974, 3194, 5168, 8362, 13530, 21892, 35422$. In the inset of Fig.~\ref{fig3}(b), we observe that as the system size $L$ increases, the second critical region persists, indicating the stability of the reentrant localization feature in the staggered system with long-range hopping. 

To better understand which eigenstates undergo reentrant localization transitions as the disorder strength $\lambda$ changes, we fix $J_{33}/J_1=2.7$ and plot colormaps of the IPR and NPR as functions of the eigenstate energy ordered index and $\lambda$ in Fig~\ref{fig4}. The colormaps reveal that some eigenstates in the central energy range transition from localized to extended states in the range $2.6<\lambda<3.6$. Furthermore, we plot the spatial distribution of the eigenstate with $m/L=0.5$ for $\lambda=0.4, 1.2, 2.1, 2.7$ and $4$. It is evident that the eigenstate is clearly localized at $\lambda=2.1$, but as $\lambda$ increases to $2.7$, the eigenstate transitions to an extended state.

The colormaps demonstrate the presence of single-particle mobility edges in this 1D system. To more clearly identify the locations of the mobility edges, we show the distributions of IPR and NPR as functions of the eigenstate index for $\lambda=0.4, 1.2, 2.1, 2.7, 4$ in Fig~\ref{fig5}(a)-Fig~\ref{fig5}(e). From Fig~\ref{fig5}, we observe that as $\lambda$ increases, single-particle mobility edges appear at $\lambda=1.2$. These mobility edges then vanish, only to re-emerge as $\lambda$ continues to increase. At higher values of $\lambda$, the mobility edges eventually disappear entirely. Similarly, the presence of single-particle mobility edges is also be evident in the system’s density of states (DOS), as shown in Fig~\ref{fig5}(f)-Fig~\ref{fig5}(j). 

It is noteworthy that in Fig~\ref{fig5}(b), the system first exhibits single-particle mobility edges, with two pairs of mobility edges observed. This behavior contrasts with systems lacking long-range hopping, where typically only one pair of single-particle mobility edges is present. Correspondingly, the DOS in Fig~\ref{fig5}(g) also displays symmetric two pairs of mobility edges. We interpret this as follows: in systems with long-range hopping, the structure can be viewed as a 1D ladder system. When there is no coupling between the two legs of the ladder (i.e., $J_{11}=J_3=0$), each leg can independently exhibit one pair of mobility edges. However, due to the difference in parameters between the two legs, the positions of the mobility edges will also differ. When $J_{11}\neq0$ and $J_3\neq0$, the coupling acts as a perturbation, which may shift the positions of the mobility edges. Nevertheless, the system retains two pairs of mobility edges. In a word that long-range hopping introduces additional pairs of mobility edges into the system, providing greater flexibility for studying the critical states in 1D systems. 

Furthermore, we investigate the critical behavior at various localization transition points. By zooming in Fig~\ref{fig4}(a) and Fig~\ref{fig4}(b), we preliminarily identify the critical disorder strengths, $\lambda_c$, associated with the localization transitions. In a narrow region around each $\lambda_c$, we calculate the $R[L, L^{'}]$ function based on Eq. \ref{R equ}. In Fig~\ref{fig6}, we present the $R[L, L^{'}]$ functions for different localization transition points. The crossing points of these $R[L, L^{'}]$ functions on the horizontal axis determine the critical disorder strengths, $\lambda_c$, for the four localization transitions indicated by the black dots in Fig~\ref{fig2}(b). The corresponding critical disorder strengths for each transition are as follows: $\lambda_{c,1}=0.9802$, $\lambda_{c,2}=2.0194$, $\lambda_{c,3}=2.599$, and $\lambda_{c,4}=3.239$. The vertical coordinates of the crossing points correspond to the critical exponent ratio $\gamma/\nu$ for the four transition points, which are given by: $\gamma_1/\nu_1=0.57\pm0.016$, $\gamma_2/\nu_2=0.46\pm0.006$, $\gamma_3/\nu_3=0.22\pm0.007$, and $\gamma_4/\nu_4=0.43\pm0.009$. Following the description in Eq. \ref{nu}, we plot $\sigma^2L^{1-\gamma/\nu}$ and $\varepsilon L^{1/\nu}$ as functions of system size, with $L=8362, 13530, 21892$ and $35422$. By adjusting the critical exponent $\nu$, we align the $\sigma^2L^{1-\gamma/\nu}$ and $\varepsilon L^{1/\nu}$ curves for different $L$, achieving optimal overlap. This approach provides an optimal estimate for the critical exponent $\nu$. The calculated values of $\nu$ for the four localization transitions are: $\nu_1=0.9$, $\nu_2=1.1$, $\nu_3=1.8$ and $\nu_4=0.9$. 

In contrast to the results obtained for systems without long-range hopping \cite{PhysRevB.105.214203}, our calculated critical exponents $\nu$ indicate that the second and third localization transition points in the model with long-range hopping do not belong to the same universality class. While the critical exponents $\nu$ for the first and fourth localization transition points are identical, their $\gamma/\nu$ values differ, suggesting that the $\gamma$ values at these two localization transition points are distinct. This implies that, in the presence of long-range hopping, the critical behaviors at the four localization transition points belong to different universality classes. 

Additionally, we calculate the critical exponents corresponding to different localization transition points using the scaling law for $NPR_L*L$. According to Eq. \ref{critical}, we know that $NPR_L*L\sim L^{\gamma/\nu}$. By fitting the linear relationship between $\log(L)$ and $log(NPR_L*L)$ using the least-squares method, we determine the critical exponent ratio $\gamma/\nu$ as shown in Fig~\ref{fig7}. The fitting yields the following critical exponent ratios for the four localization transition points: $\gamma_1/\nu_1=0.5735\pm 8e-05$, $\gamma_2/\nu_2=0.4607\pm 1.4e-04$, $\gamma_3/\nu_3=0.2243\pm 2.3e-05$ and $\gamma_4/\nu_4=0.4309\pm 1e-04$. These results are consistent with those obtained previously from the vertical coordinates of the $R[L, L’]$ function. Furthermore, from Eq. \ref{critical}, we also know that $\sigma_L\sim L^{-\beta/\nu}$. Similarly, by applying the least squares method to fit the linear relationship between $\log(L)$ and $\log(\sigma_L)$, we determine the corresponding critical exponent $\beta/\nu$ for the four localization transition points as follows: $\beta_1/\nu_1=0.2133\pm 2e-05$, $\beta_2/\nu_2=0.2696\pm 3.5e-05$, $\beta_3/\nu_3=0.3879\pm 6e-06$ and $\beta_4/\nu_4=0.2846\pm 2.6e-05$. The critical exponents we obtained should satisfy the hyperscaling law \cite{hashimoto1992finite} as
\begin{equation}
	\frac{2\beta}{\nu}+\frac{\gamma}{\nu}=1. 
	\label{hyperscale}
\end{equation}
We confirm that the critical exponents obtained in this study satisfy the hyperscaling relation.

\section{Uniform disorder}\label{se4}

\begin{figure}
    \centering
    \includegraphics[width=0.5\textwidth]{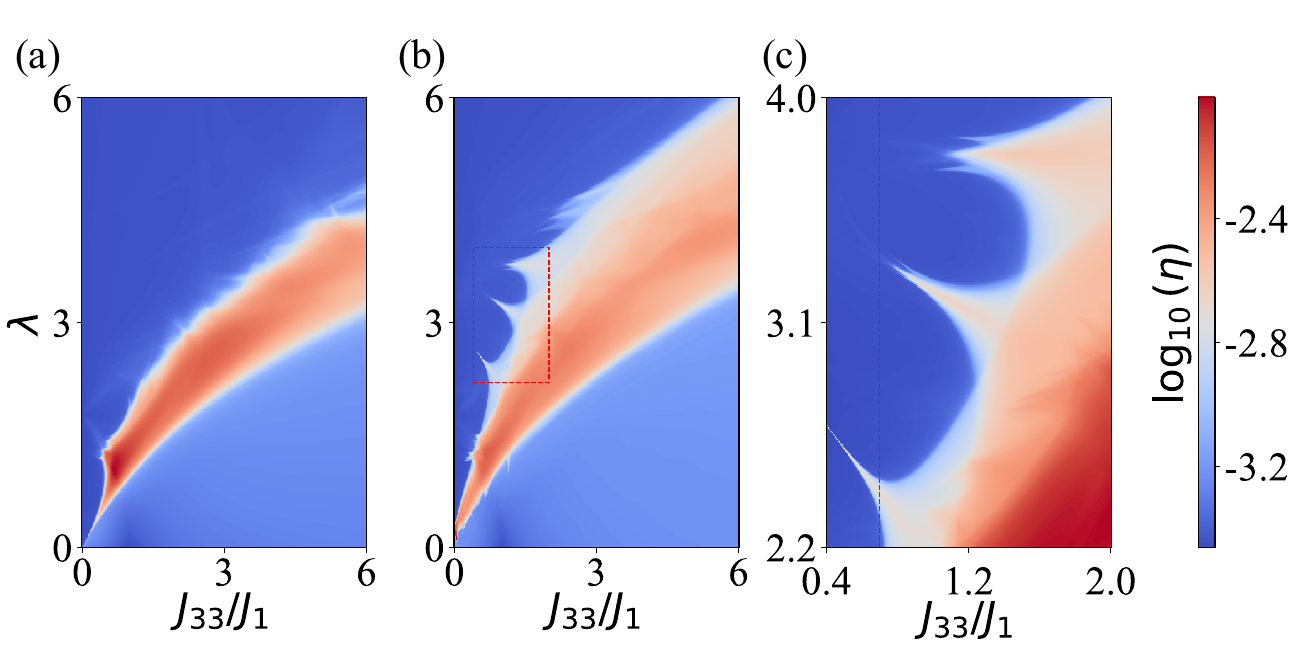}
    \caption{$\eta$ phase diagrams of the system in $\lambda$ and $J_{33}/J_{1}$ plane with (a) $J_{11}=J_3=0$, (b) $J_{11}=J_3=0.1$, the system size $L=3194$, $\beta=\frac{\sqrt{7}-1}{2}$ and $\lambda_A=\lambda_B$. (c) The zoomed-in version of the portion enclosed by the dashed box in (b). The color represents different values of $\log_{10}(\eta)$. The dashed line in (c) is located at $J_{33}/J_1=0.7$. }
    \label{fig8}
\end{figure}

\begin{figure}
     \centering
     \includegraphics[width=0.48\textwidth]{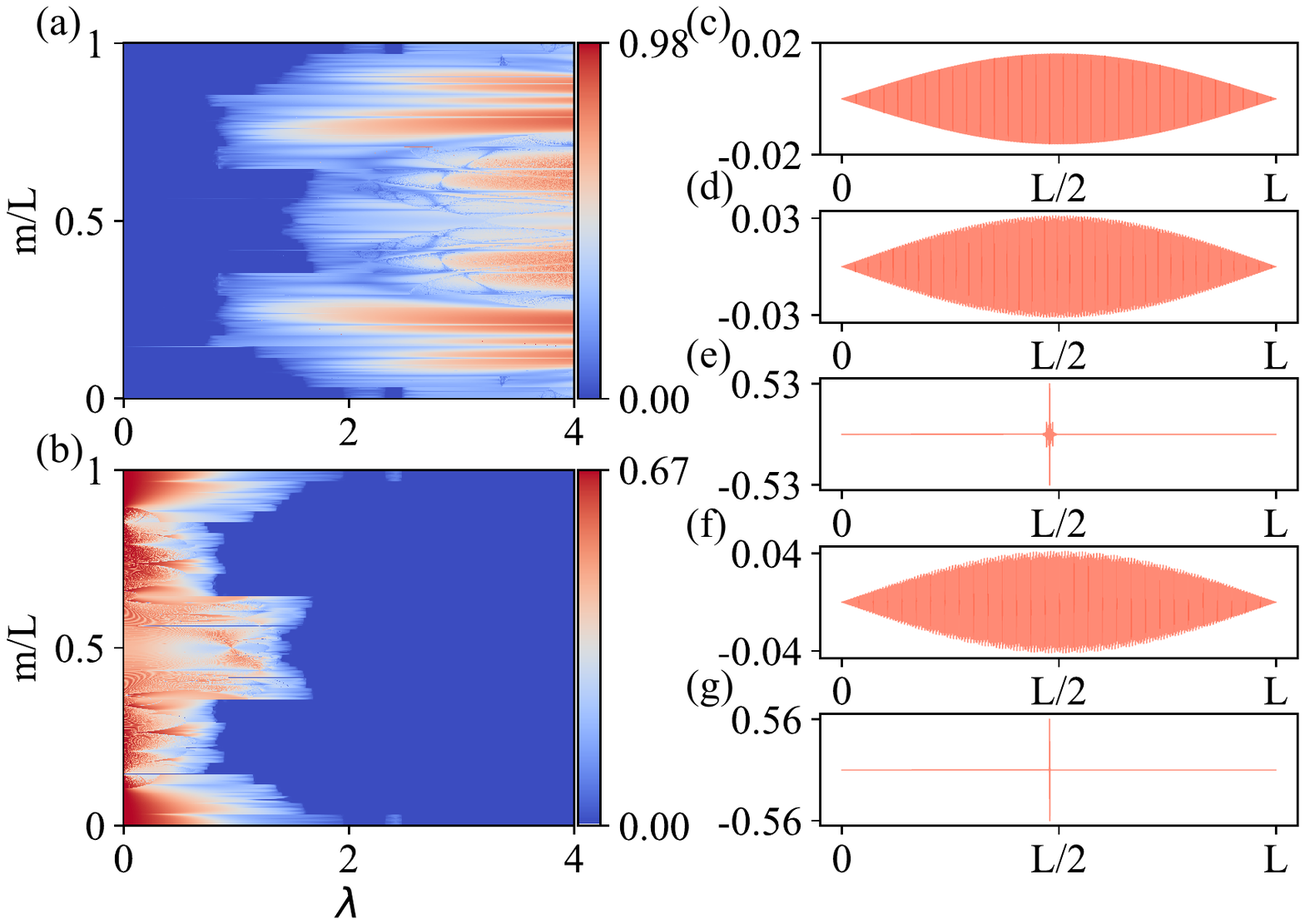}
     \caption{(a) The IPR and (b) NPR associated with the eigenstate indices as a function $\lambda$ for $J_{33}/J_1=0.7$, respectively. (c) to (g) illustrate the spatial distribution of eigenstates across lattice sites for $\lambda=0.4, 1.2, 2.2, 2.41, 4$, with eigenstate index $m/L=0$ and $L=13530$. }
     \label{fig9}
\end{figure}

\begin{figure}
     \raggedright
     \includegraphics[width=0.45\textwidth]{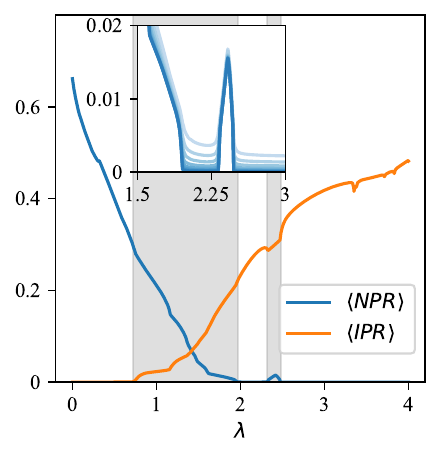}
     \caption{The average IPR and NPR over eigenstates with $m/L\in [0, 1/6]$ for $J_{33}/J_1=0.7$, for the case of uniform disorder and system size $L=13530$. The shaded regions represent the critical area, where both localized and extended states coexist. The inset shows the average NPR for $L=1220,  1974,  3194,  5168,  8362, 13530, 21892,$ and $35422$ with the color intensity ranging from light to dark. }
     \label{fig10}
\end{figure}

\begin{figure*}
     \centering
     \includegraphics[width=\textwidth]{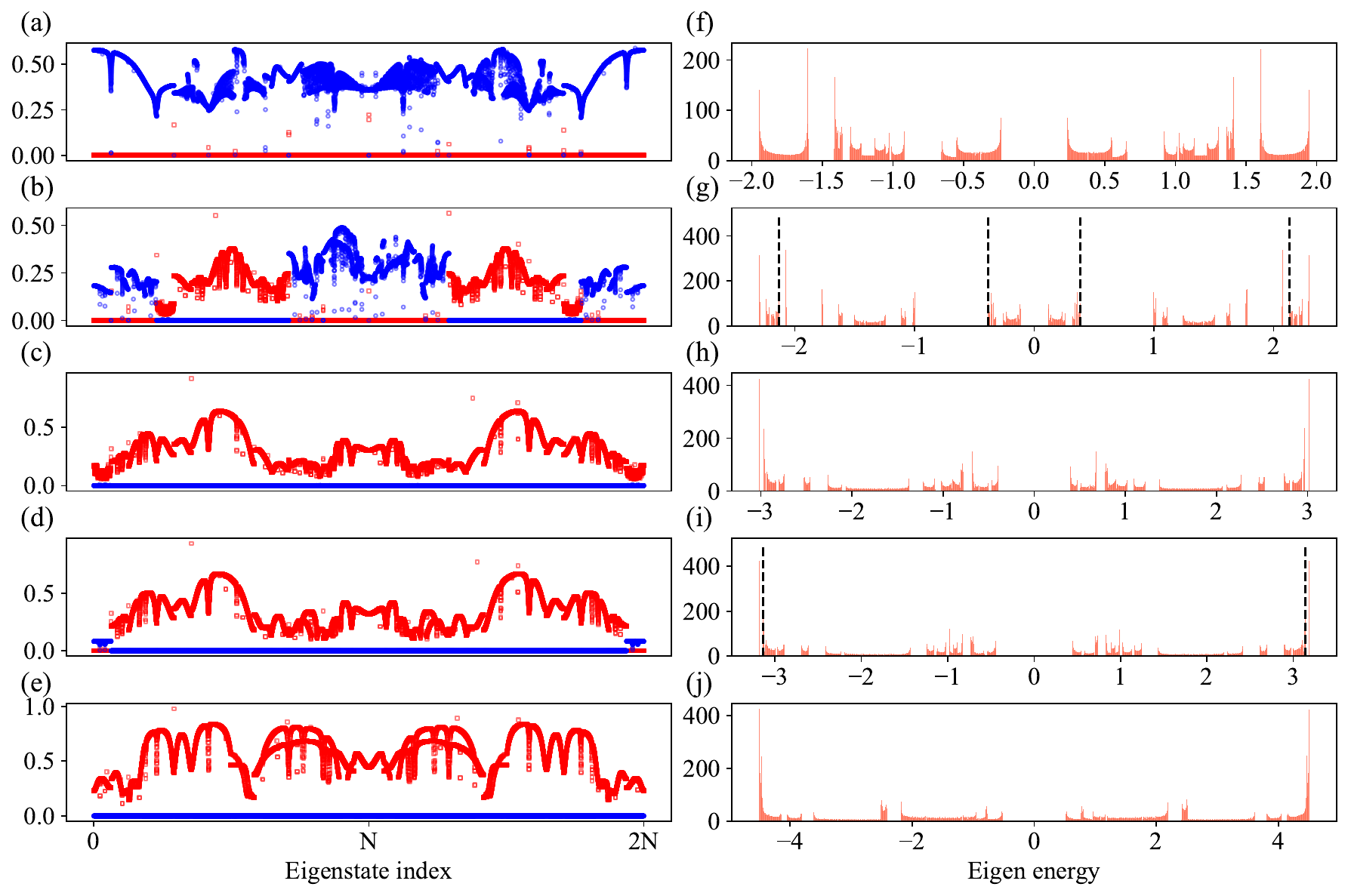}
     \caption{(a)–(e) show the distributions of IPR (red) and NPR (blue) as functions of the eigenstate index for $\lambda=0.4$, $1.2$, $2.2$, $2.41$ and $4$, respectively. (f)–(j) display the DOS as a function of eigenenergy for the same $\lambda$ values, with the black dashed lines indicating the locations of single-particle mobility edges. }
     \label{fig11}
\end{figure*}

\begin{figure}
     \centering
     \includegraphics[width=0.45\textwidth]{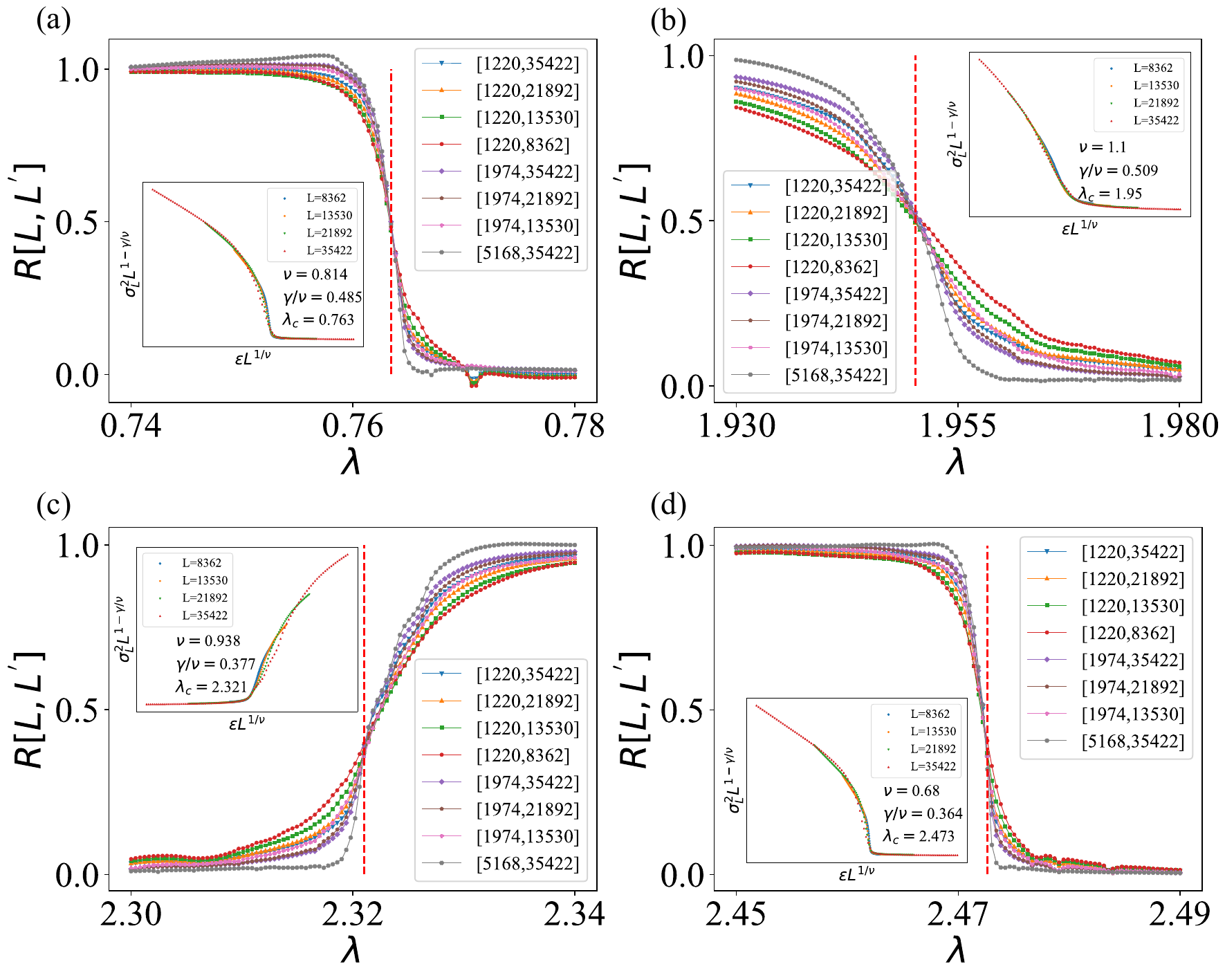}
     \caption{(a)–(d) depict the values of the function $R[L,\ L^{'}]$ near the first, second, third, and fourth critical quasi-periodic disorder potential strength, with $J_{33}/J_1=0.7$ and $\lambda_A=\lambda_B$. In each panel, the inset presents the $\sigma^2$ of the best fitted critical exponent $\nu$ near the respective critical quasi-periodic disorder potential strength. The coordinates of the function $R[L,\ L^{'}]$ crossing point is denoted as $(\lambda_c, \gamma/\nu)$. For the first critical point, we average the NPR of eigenstates with $m/L \in [0.167, 0.177]$. For the second, third, and fourth critical points, we average the NPR of eigenstates with $m/L \in [0, 0.1]$. }
     \label{fig12}
\end{figure}

\begin{figure}
     \centering
     \includegraphics[width=0.48\textwidth]{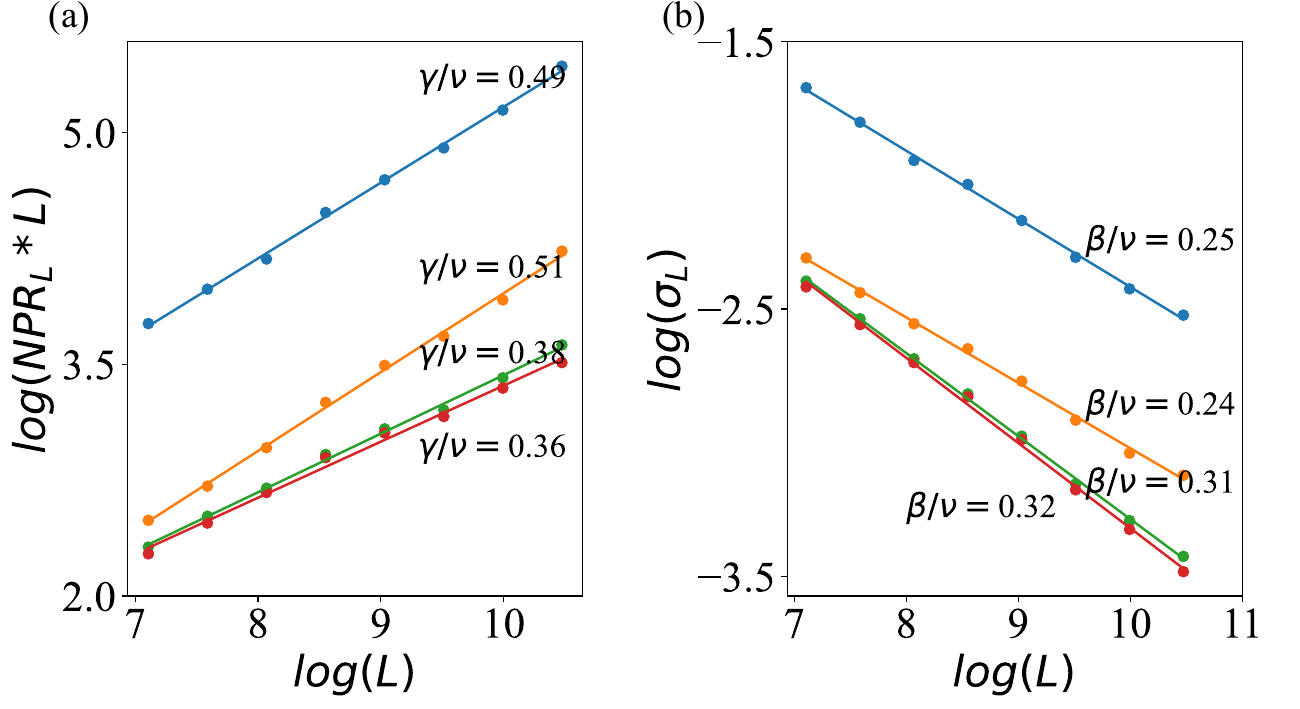}
     \caption{The critical exponent ratio (a) $\gamma/\nu$ and (b) $\gamma/\beta$ for the four critical quasi-potential points. These ratios are obtained by fitting $\log(NPR_L*L)$ against $\log(L)$. The system sizes $L$ used for the fitting were: $1220, 1974, 5168, 8362, 13530, 21892$, and $35422$. }
     \label{fig13}
\end{figure}

Previous studies have shown that reentrant localization transitions occur exclusively in 1D systems with staggered disorder or with a large detuning between intra-cell and inter-cell hopping in the presence of uniform disorder \cite{gong2021comment}. In contrast, for 1D systems with uniform disorder and small detuning between intra-cell and inter-cell hopping, only single-particle mobility edges have been reported. In this work, we investigate the localization properties of a 1D system with uniform disorder in the presence of long-range hopping. Our results reveal that, when long-range hopping is introduced, the system exhibits reentrant localization transitions, a phenomenon that was previously not observed in uniform disorder systems without long-range hopping under small detuning.

Initially, we consider the case where $J_{11} = J_3 = 0$. In this scenario, the system can be treated as a 1D ladder model with no coupling between the two legs of the ladder. Under these conditions, we calculate the system's $\eta$-phase diagram, as shown in Fig~\ref{fig8}(a). The diagram clearly reveals a critical transition region, indicating that as the disorder strength increases, the system transitions from an extended state to a regime where some eigenstates become localized. At this stage, the system exhibits single-particle mobility edges. As the disorder strength is further increased, the system eventually fully transitions into a localized state. These observations are consistent with previous studies on systems with uniform disorder. 

Next, we introduce coupling between the two legs of the ladder structure by setting the inter-leg hopping parameters to $J_{11}=J_3=0.1$. Under these conditions, we recalculate the $\eta$-phase diagram. Compared to Fig~\ref{fig8}(a), the updated phase diagram reveals additional spike-like structures, highlighted by the dashed box in Fig~\ref{fig8}(b). To investigate this further, we zoom in on the region within the dashed box, as shown in Fig~\ref{fig8}(c). In this magnified view, we observe that at one of the spikes, as the disorder strength $\lambda$ increases, the system transitions from an extended state to a critical state, and eventually to a fully localized state. Notably, as $\lambda$ continues to increase, some localized eigenstates revert to extended states within the spike. Eventually, when $\lambda$ becomes sufficiently large, all eigenstates become fully localized. 

We select the parameter $J_{33}/J_1 = 0.7$, as shown in Fig~\ref{fig8}(c), and calculate distributions of the system's IPR and NPR as functions of disorder strength $\lambda$ and eigenstate index. Fig~\ref{fig9}(a) and ~\ref{fig9}(b) show that, as $\lambda$ increases, most eigenstates transition from extended to localized states. However, for eigenstates near the lowest and highest eigenenergy, a reentrant localization transition occurs. To further illustrate this phenomenon, Fig~\ref{fig9}(c)-Fig~\ref{fig9}(g) depict the spatial distribution of a specific eigenstate indexed by $m/L=0$ for $\lambda$ values of $0.4$, $1.2$, $2.2$, $2.41$, and $4$, respectively. At $\lambda = 2.2$, the eigenstate transitions into a localized state. As $\lambda$ increases to $2.41$, the eigenstate reverts to an extended state, before localizing again as $\lambda$ increases further. Notably, comparing the delocalization observed here (Fig~\ref{fig9}(f)) with that observed under staggered disorder conditions (Fig~\ref{fig4}(f)), we find that the eigenstates in Fig~\ref{fig9}(f) exhibit a more extended spatial distribution. 

To rule out the finite-size effects, we calculate the average IPR and NPR over eigenstates with $m/L\in [0, 1/6]$ as functions of disorder strength $\lambda$, and plot the results in Fig~\ref{fig10}. In the second shaded region, a characteristic feature of reentrant localization emerges. The inset of Fig~\ref{fig10} provides a magnified view of this region, showing the results for different system sizes: $L=1220,  1974,  3194,  5168,  8362, 13530, 21892,$ and $35422$ with colors ranging from light to dark. For all system sizes, the reentrant localization feature persists, confirming that the observed localization transition is not influenced by finite-size effects.

To further explore how the mobility edges evolve with increasing disorder strength $\lambda$, we plot the distributions of IPR and NPR as functions of the eigenstate index in Fig~\ref{fig11}(a)-Fig~\ref{fig11}(e) for $\lambda=0.4$, $1.2$, $2.2$, $2.41$ and $4$, respectively. At $\lambda=0.4$ (Fig~\ref{fig11}(b)), the system exhibits mobility edges, which appear in pairs and separate localized and extended states within the spectrum, indicating the coexistence of both extended state and localized state of eigenstates. As $\lambda$ increases to $2.2$ (Fig~\ref{fig11}(c)), the mobility edges disappear, and all eigenstates become localized, signaling a complete transition to a fully localized phase. Interestingly, at $\lambda=2.4$ (Fig~\ref{fig11}(d)), a subset of eigenstates transitions back to extended states, leading to the reappearance of mobility edges. In this case, the extended states are located near the highest and lowest eigenenergy, in contrast to the staggered disorder scenario, where the extended states typically emerge in the middle of the eigenenergy spectrum. Finally, as $\lambda$ increases further to $4$ (Fig~\ref{fig11}(e)), all eigenstates become localized again. Corresponding DOS plots are shown in Fig~\ref{fig11}(f)-Fig~\ref{fig11}(j), which clearly reflect the changes in the mobility edges as $\lambda$ varies. The appearance, disappearance, and reappearance of extended states align with the observations from the IPR and NPR distributions.

In Fig~\ref{fig12}, we calculate the critical exponents at different localization transition points. Using the parameters marked by the dashed line in Fig~\ref{fig8}(c), specifically $J_{33}/J_1=0.7$, we evaluate the $R$-function in a small neighbourhood around the critical disorder strengths $\lambda_c$. As discussed previously, the horizontal coordinate of the intersection points of the $R$-function provides the critical disorder strength $\lambda_c$ for each localization transition, while the vertical coordinate corresponds to the ratio of critical exponents $\gamma/\nu$. From the analysis of the $R$-function results in Fig~\ref{fig12}, we determine the critical disorder strengths for the four localization transitions as $\lambda_{c, 1}=0.7634,\ \lambda_{c, 2}=1.9502,\ \lambda_{c, 3}=2.3210$ and $\lambda_{c, 4}=2.4726$, with the corresponding critical exponent ratios $\gamma_1/\nu_1=0.4849\pm0.0092,\ \gamma_2/\nu_2=0.5088\pm0.0113,\ \gamma_3/\nu_3=0.3771\pm0.012$ and $\gamma_4/\nu_4=0.3645\pm0.0221$. Based on the description of Eq. \ref{nu}, we plot the relationship $\sigma^2L^{1-\gamma/\nu}(\varepsilon L^{1/\nu})$ in the insets of Fig~\ref{fig12}. For systems of different sizes, the equation should follow the same functional relationship when calculated using the optimally fitted $\nu$. Accordingly, we perform calculations for four systems with sizes $L=8362, 13530, 21892$ and $35422$. The fitting process yields the following optimal critical exponents: $\nu_1=0.814,\ \nu_2=1.10,\ \nu_3=0.938,$ and $\nu_4=0.680$. These results are consistent with those obtained for the staggered disorder system, confirming that each of the four localization transition points exhibits distinct critical exponents. 

Similarly, we can use the scaling law of $NPR_L*L$ to fit the critical exponents, as shown in Fig~\ref{fig13}. According to Eq.~\ref{critical}, by fitting the slope of $\log(L)$ and $\log(NPR_L*L)$, we obtain the critical exponent ratios $\gamma/\nu$ at the four localization transition points: $\gamma_1/\nu_1=0.4907\pm 7.6e-05$,\ $\gamma_2/\nu_2=0.5106\pm 1e-04$,\ $\gamma_3/\nu_3=0.3792\pm 7e-05$ and $\gamma_4/\nu_4=0.3629\pm 1.48e-04$. These results are consistent with those obtained using the $R$-function approach discussed above. Furthermore, by fitting the slope of $\log(L)$ versus $\log(\sigma_L)$, we derive the critical exponent ratios $\beta/\nu$, which at the corresponding transition points are $\beta_1/\nu_1=0.2547\pm 1.9e-05$,\ $\beta_2/\nu_2=0.2447\pm 2.5e-05$,\ $\beta_3/\nu_3=0.3104\pm 1.7e-05$ and $\beta_4/\nu_4=0.3185\pm 3.7e-05$. Additionally, the obtained critical exponent ratios satisfy the hyperscaling law described by Eq.~\ref{hyperscale}, $\frac{2\beta}{\nu}+\frac{\gamma}{\nu}=1$.

\section{Influence of Inter-leg hopping}\label{se5}
\begin{figure}
    \centering
    \includegraphics[width=0.5\textwidth]{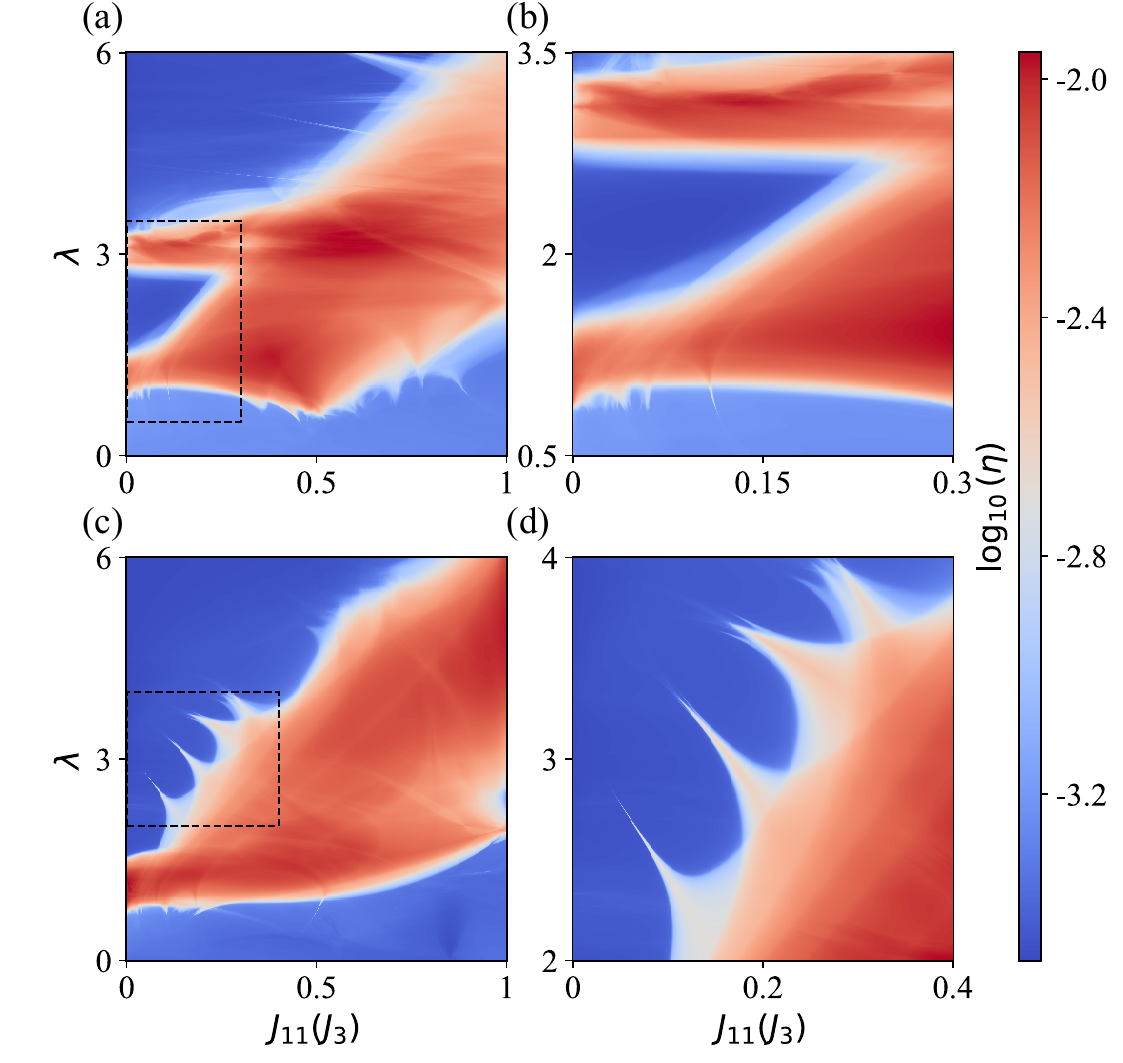}
    \caption{The $\eta$ phase diagrams of the system in the disorder strength $\lambda$ and inter-leg hopping $J_{11}(J_3)$ plane with (a) $J_{33}/J_1=2.7$ under staggered disorder and (c) $J_{33}/J_1=0.7$ under uniform disorder. (b) and (d) present magnified views of the regions outlined by the dashed boxes in (a) and (c), respectively. All the phase diagrams are calculated under a system size $L=3194$. }
    \label{fig14}
\end{figure}

\begin{figure}
    \centering
    \includegraphics[width=0.5\textwidth]{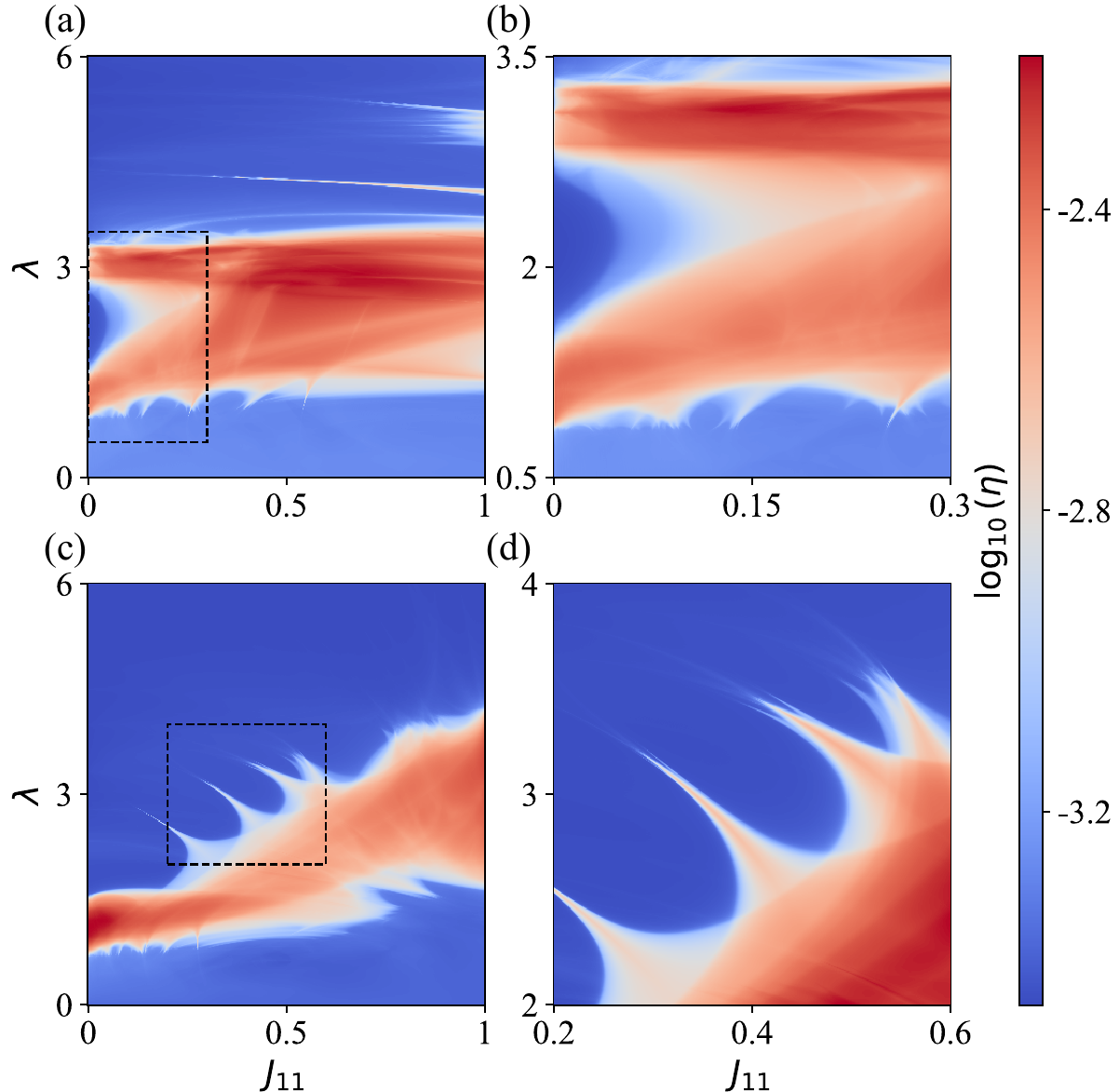}
    \caption{The $\eta$ phase diagrams of the system in the disorder strength $\lambda$ and inter-leg hopping $J_{11}$ plane for (a) $J_{33}/J_1=2.7$ under staggered disorder and (c) $J_{33}/J_1=0.7$ under uniform disorder. (b) and (d) present magnified views of the regions outlined by the dashed boxes in (a) and (c), respectively. All the phase diagrams are calculated under a system size $L=3194$ and $J_3=0$. }
    \label{fig15}
\end{figure}

In this section, we separately examine the influence of inter-leg hopping on reentrant localization in the presence of staggered and uniform disorder. As illustrated in the schematic diagram in Fig~\ref{fig0}(b), the inter-leg hopping is defined by the parameters $J_{11}$ and $J_3$. Using the parameters marked by dashed lines in Fig~\ref{fig2}(c) and Fig~\ref{fig8}(c), we investigate the effect of varying $J_{11}$ and $J_3$, while maintaining $J_{11}=J_3$. The corresponding $\eta$-phase diagrams are shown in Fig~\ref{fig14}.

Fig~\ref{fig14}(a) and Fig~\ref{fig14}(b) illustrate the case for staggered disorder. As the inter-leg hopping strength increases, the previously observed reentrant localization feature gradually diminishes, eventually disappearing entirely at $J_{11}=J_3=0.207$. This behavior can be attributed to the competition between dimerized hopping and staggered disorder within each individual leg of the ladder. The inter-leg hopping weakens the effective dimerization strength, leading to the suppression of reentrant localization as inter-leg hopping increases.

For the case of uniform disorder, as shown in Fig~\ref{fig14}(c) and Fig~\ref{fig14}(d), reentrant localization does not initially appear at $J_{11}=J_3=0$. However, as the inter-leg hopping strength increases, the $\eta$-phase diagram begins to exhibit a “spiky” structure similar to that observed in Fig~\ref{fig8}(c). These spikes indicate the emergence of reentrant localization transitions. With further increases in the inter-leg hopping strength, these spikes persist at various $J_{11}=J_3$ values but eventually vanish as the inter-leg hopping $J_{11}(J_3)>0.35$. This demonstrates that, for uniform disorder, the appearance of reentrant localization requires a minimum of inter-leg hopping strength.

In Fig~\ref{fig15}, we examine the case where $J_{3}=0$, simplifying the ladder structure in Fig~\ref{fig0} to one without slanted hopping. By calculating the $\eta$-phase diagrams, we also observe distinct behaviors under staggered and uniform disorder. For staggered disorder, the inter-leg hopping $J_{11}$ progressively diminishes the originally observed reentrant localization phenomenon, consistent with our earlier findings that inter-leg hopping disrupts the competition between dimerized hopping and staggered disorder, which is crucial for reentrant localization. In contrast, for uniform disorder, the emergence of reentrant localization requires a nonzero inter-leg hopping $J_{11}$. However, once $J_{11}$ exceeds $0.51$, the reentrant localization phenomenon disappears. This suggests that while a moderate strength of inter-leg hopping is necessary to induce reentrant localization under uniform disorder, excessive inter-leg hopping ultimately suppresses this behavior. These results highlight the crucial role of inter-leg hopping in shaping the localization properties of the system under different disorder conditions.

\section{Conclusion}\label{se6}

In conclusion, we have investigated the phenomenon of reentrant localization, which has previously been observed in systems without long range hopping. While long-range hopping is generally expected to disrupt reentrant localization, our results show that under specific parameter conditions, varying the strength of long-range hopping can induce a reentrant localization phase diagram, similar to that found in staggered SSH systems. We conduct a detailed analysis of the observed reentrant localization phenomena, including the spatial distribution of eigenstates across lattice sites and the presence of multiple mobility edges. These features confirm that the system undergoes a genuine reentrant localization transition. Additionally, our calculation of critical exponents reveals that long-range hopping gives rise to four distinct localization transition points, and each characterized by unique critical exponents. Previous studies have suggested that systems with uniform disorder and small inter-intra detuning do not exhibit reentrant localization. In contrast, our work demonstrates that long-range hopping can induce reentrant localization even in the presence of uniform disorder. Furthermore, we observe that the reemergent extended states are located near the highest and lowest eigenenergy, which contrasts with the typical behavior observed in staggered disorder systems where extended states generally appear near the middle of the eigenenergy spectrum. Finally, we explore the critical exponents associated with these transitions. Our results indicate that the inclusion of long-range hopping leads to distinct critical exponents at all four transition points, highlighting the complex and unique nature of localization phenomena in systems with long-range interactions. 

\begin{acknowledgments}
The authors would like to thanks Yuhang Lu for his helpful discussion. This work is supported by the National Natural Science Foundation of China under Grants No. 11974205, and No. 61727801, the Key Research and Development Program of Guangdong province (2018B030325002), and National Natural Science Foundation of China under Grant 62131002. 
\end{acknowledgments}

\bibliography{apssamp}

\end{document}